# Could one make a diamond-based quantum computer?

A Marshall Stoneham, A H Harker, Gavin W Morley
London Centre for Nanotechnology, University College London, London WC1E 6BT

**Abstract**     We assess routes to a diamond-based quantum computer, where we specifically look towards scalable devices, with at least 10 linked quantum gates. Such a computer should satisfy the deVincenzo rules and might be used at convenient temperatures. The specific examples we examine are based on the optical control of electron spins. For some such devices, nuclear spins give additional advantages. Since there have already been demonstrations of basic initialisation and readout, our emphasis is on routes to two-qubit quantum gate operations and the linking of perhaps 10-20 such gates. We analyse the dopant properties necessary, especially centres containing N and P, and give results using simple scoping calculations for the key interactions determining gate performance. Our conclusions are cautiously optimistic: it may be possible to develop a useful quantum information processor that works above cryogenic temperatures.

## 1.     Introduction

In his Newton Medal talk, London 2008 Anton Zeilinger said:  "We have to find ways to build quantum computers in the solid state at room temperature - that's the challenge." Indeed it is a challenge. But a recent analysis (Stoneham 2009) gave cautious grounds for optimism, with diamond as a key material.

Quantum computers differ from classical computers in that they exploit entanglement, the correlations between *local* measurements on two particles. Such a computer, successfully implemented, could carry out some potentially important classes of calculation well beyond the capabilities of classical devices. There is no such computer today, and indeed doubts have been expressed about the possibilities of success. It is prudent therefore to look at the current successes and future challenges to assess the feasibility of a quantum computer based on diamond.

Broadly speaking, three classes of quantum computer have been suggested that exploit favourable properties of diamond. In Class 1, the demonstrated special qualities of an $NV^-$ centre (discussed below) are exploited, usually making use of the $NV^-$ electron spin to manipulate nearby carbon or nitrogen nuclear spins (e.g., Jelezko and Wrachtrup 2004, Wrachtup and Jelezko 2006, Greentree et al 2006). Class 2 follows the ideas of Stoneham, Fisher and Greenland (2003, to be referred to as SFG), requiring two dopant or defect species. The electron spins of one species (perhaps substitutional N, or possibly $NV^-$) provide qubits, and these are manipulated by electronically-excited control species (perhaps substitutional P). In Class 3, quantum entanglement is created between *remote* systems (Cabrillo et al 1999), such as $NV^-$ centres in distinct diamonds, by a measurement involving "single shot" excitation (Benjamin et al 2009) that leaves the two systems entangled, and available for various quantum compoutation strategies, such as measurement-based quantum computing.

The present paper looks primarily at Class 2 approaches, based on the optical control of localised defects in diamond. Two important questions concern decoherence times and scalability. Will quantum entanglement last long enough for useful calculations? Can one link enough qubits together successfully to make a useful computer? As discussed below, there are many papers on Class 1 approaches, with demonstrations of quantum operations and simple gates. However, it is not at all obvious how Class 1 can be scaled up, despite some very ingenious experimental work, like the creation of close pairs of $NV^-$ centres by implantation of nitrogen molecules. There has been informal talk of nanoelectrodes, but these seem unlikely to be practical, even if they can be fabricated, if only from issues such as charge fluctuations and noise. Class 3 show promise, but will be discussed here only to define some of the condensed matter physics factors that may prove limiting.

For any practical quantum computer, major challenges include issues of fabrication, and the linkage of the quantum device to the classical silicon-based and photonic devices that will be used to run it. Integration of the whole system may prove especially hard, given the length scale and demands of the active components. Ideally, the system should be one that could be created in a current or near-future fabrication plant. The Class 2 (SFG) approach is of this type, though the complexity of system integration should not be underestimated. The specifically materials issues have been discussed by us previously (Stoneham 2005, 2008), and will not be repeated here, where we shall concentrate primarily on the species that might be exploited in an SFG scheme. We combine simple scoping calculations with available experimental information to analyse what might be feasible.

## 2. Basic Issues of Quantum Information Processing

Both Class 2 and Class 3 approaches appear to have the potential to be scaled to reasonably large computer sizes, to be compatible with current silicon microelectronic and laser technologies, and to work well above cryogenic temperatures. For Class 1, scalability is more questionable. But any quantum computer can succeed only if it can meet a whole series of other challenges. One set of challenges is summarised in the DiVincenzo criteria (DiVincenzo and Loss 1998). In describing these, we assume the qubits are electron spins, unless otherwise stated. Clearly, there are ways to take advantage of nuclear spins, especially as quantum information stores, but analysing such routes is not part of the present paper.

The DiVincenzo rules, briefly summarised, are these:

**(i)** *Is there a well defined Hilbert space to represent the quantum information?* The qubit spins do define a suitable space. There could be problems at high temperatures, e.g., if spins ionise thermally, but room temperature is cool enough for diamond dopants.

**(ii)** *Initializing qubit states.* Several approaches are possible, some exploiting selection rules in optical or microwave transitions; these have already been demonstrated for $NV^-$ centres. Alternative methods, such as spintronic approaches in which the system is flushed with spin-polarised carriers, are possible but probably unnecessary. Polarized electronic spin qubits can be used to polarize coupled nuclear spin qubits (Morley et al 2007) Brute force methods needing thermalisation in an applied field need Zeeman energies much larger than thermal energies, $g\beta H \gg kT$, and this can be difficult to meet except at low temperatures (Morley et al 2008)

**(iii)** *Manipulating quantum information.* Manipulations usually need faster transitions, so as to give effective switching at faster clock speeds. Most computer architectures demand some way to select one quantum gate rather than another, and this can put limits on acceptable optical linewidths.

**(iv)** *Avoid decoherence for long enough to compute.* Entanglement is the resource that could make quantum computing worthwhile, and describes the correlations between local measurements on the qubit spins that might be described informally as their "quantum dance." The enemy of entanglement is *decoherence*, just as friction is the enemy of mechanical computers. One obvious decoherence mechanism is spin relaxation, which degrades both entanglement and quantum information stored as qubits. Spin-spin relaxation, from the interactions of qubit spins with any other spins - nuclear or electron - can be minimised in various ways, e.g., by eliminating nuclear spins in an isotopically pure sample (Tyryshkin et al 2006). In the SFG scheme, spins will be in modest-sized patches, perhaps a few 100 nm across, and containing only perhaps a few hundred spins. This already makes flip-flop processes (from $S_{1+}S_{2-}$ dipole dipole terms) less favourable as energy conservation by the spin-spin interaction reservoir is more limited. Further suppression is possible with controlled inhomogeneity, e.g., through the effect of strain on g factors (Stoneham 1975 section 12.4.2). Spin-lattice relaxation is especially slow in diamond, partly because of the small spin-orbit coupling and partly because of the high sound velocity [the direct process rate $\sim v^{-5}$, Raman (two phonon) rate $\sim v^{-10}$ (Stoneham 1975 section 14.2.2). Even at room temperature, $N_s$ has a relaxation time of order 1 ms.

**(v)** *Readout of the quantum information.* This needs to be done quickly, before quantum information is lost. There are many options for readout of single spins, some demonstrated for the NV$^-$ centre in diamond (e.g., Jelezko and Wrachtrup 2004, Popa et al 2004, Wrachtrup and Jelezko 2006). We aim to discuss alternative strategies to sequential readout in a separate paper.

**(vi)** *Scalability.* Can one combine the manipulations of individual qubits into a system that manipulates usefully large numbers of qubits? There is little point in doing a quantum calculation unless it can achieve more than a calculation on existing classical computers. Using quantum methods to factorise 15, or search a directory of 4 items impressively demonstrate principles, but hardly justify quantum methods, any more than they would have justified a mechanical computer in the 19$^{th}$ century. For present purposes, we assume that a quantum computer with say 20 qubits or 20 gates has promise. One would imagine such "mini-computers" linked into useful combinations, a non-trivial achievement. But a dozen linked mini-computers - say 200-250 qubits - would be a useful device, at least if error correction were not needed. System integration is certainly a major challenge, and we shall discuss some of the possible strategies.

### 3. Diamond Quantum Information Processing: present status

Some quantum operations have already been demonstrated at room temperature in diamond (e.g., Hanson et al 2006a), but progress towards an integrated system with tens or hundreds of linked quantum gates is a far harder challenge. Diamond is what one might call silicon-compatible. Of course, using silicon directly, especially if fabrication were feasible in a fabrication plant reasonably similar to the ones that exist today, would make practical QIP more likely.

As already noted (Stoneham 2008), diamond has many good features. There is a wealth of spectroscopy (Bridges et al 1990, Clark et al 1992 in Field 1992), optical and spin resonance. The spectra show many very sharp spectral lines, including a few that are sharp at room temperature. The low spin-orbit coupling and high sound velocity minimise spin-lattice relaxation. Transition energies of many defects and dopants lie in convenient ranges. Regrettably, there are also bad features. It is not easy to prepare the defect species you want. The choices of donors are very limited. Since it is not practical to stress diamond enough to split acceptor levels by more than largest phonon energies, the spin lattice relaxation of acceptors like B makes them hard to use effectively.

There are already many experiments that identify quantum operations in diamond (e.g., Wrachtrup and Jelezko 2006). Thus there has been optical initialisation (e.g., van Oort and Glasbeek 1989, Mayer Alegre et al 2007) and readout (e.g., Jelezko and Wrachtrup 2004). There have been manipulations of a single spin, equivalent to a so-called A gate (e.g., Hanson et al 2006b). Quantum operations have been demonstrated using the $NV^-$ electron spin and $^{15}N$ nuclear spin (Kennedy et al 2002), using the $NV^-$ electron spin and $^{13}C$ nuclear spin (e.g., Wrachtrup et al 2001; Jelezko et al 2004a; Popa et al 2004), and involving $NV^-$ and nearby $N_s$ (e.g., Glasbeek and van Oort 1990, Hanson et al 2007a,b). There will surely be more examples by the time this paper appears.

## 4    System needs for diamond-based quantum information processing: Qubit and Control Species for the SFG approach:

In devising the SFG gate, the aim was to have a system in which there were no ultra-nano electrodes, no need for rapid switching of applied fields, and no need for exact placements of atoms or molecules. Its basic ingredients are as follows. First, there are dopants that act as qubits that store quantum information. As such, they must be stable at the operating temperature and should have long decoherence times. If the qubits are electron spins in diamond, likely qubits might be substitutional N or $NV^-$ centres. In principle, one could arrange to pass quantum information from an electron spin to a longer-lived nuclear spin, but that introduces extra complexity and opportunities for error. Secondly, there is a distinct dopant species, the control species, used to manipulate the qubits. Ideally, the qubits and control interact negligibly in their electronic ground states. When a control dopant is excited, the more extended excited state wavefunction allows it to interact with qubits, leading to an entangling interaction between such qubits. A possible control species in diamond is substitutional P.

*FIGURE 1: The Stoneham Fisher Greenland (SFG) approach. Quantum gates exploit optical control of electron spins. Coupling between qubits (green dots) is only significant when the control dopants (red dots) are excited. The laser source can focus only to about a square wavelength, far larger than spacings of a few nm needed for entanglement. Selection of a particular gate (control plus qubits) exploits inhomogeneous broadening: different wavelengths excite different gates in a random system. Thus in (a) red entangles one group, whereas (b) green entangles another group, and leaves the first group unaffected; (c) after two excitations, two gates have made controlled changes in entanglements*

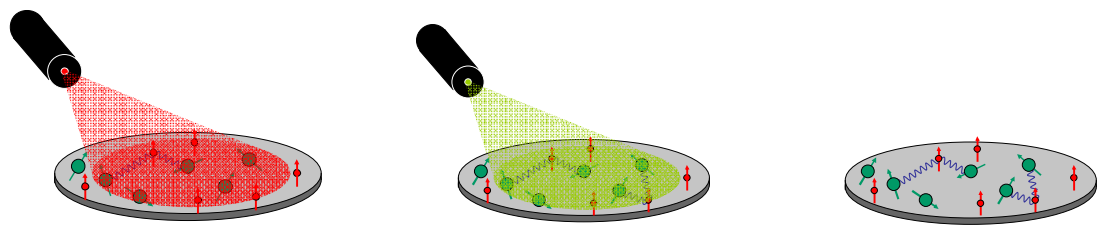

The spacings between these various qubit and control species must allow a reasonable magnitude of the interaction when the control is excited. This means (see below) spacings of a few nm. Such spacings are some two orders of magnitude less than the wavelengths of light used to excite the controls. Clearly, it is impossible to focus light down on just a single gate, i.e., a single control plus the nearby qubits whose entanglement is to be manipulated. However,

*inhomogeneity* is an important resource. The excitation energy of any particular control/qubit group will depend on the precise distances and orientations of the qubit dopants relative to the control dopant. Moreover, the local strain, the local electric fields, and - if the surface is nearby - on the presence of surface steps, all will lead to transition energies that vary from one gate to another. What matters is the excitation energy for a particular gate (control plus relevant qubits) and not whether the inhomogeneity mechansim operates primarily on the qubit or the control. In short, randomness is *desirable*, and is exploited in the SFG approach. Gate selection is achieved partly by spatial discrimination (optics, to 1-2 μm say), and partly by spectral discrimination, using the fact that key frequencies vary from one control site to the next. If need be, there are ways to enhance this inhomogeneity.

The qubits, which store quantum information, will normally have spin ½, like substitutional N. Spin 1 systems might be used if the ground state is S=1 (like $NV^-$) but most systems with even numbers of electrons have spin zero ground states (like $PV^-$). Spin systems with S=1 or larger often have faster decoherence from δS=1 spin lattice relaxation transitions and bigger spin-spin relaxation rates. The qubits can be relatively compact (like $NV^-$ or substitutional N). Acceptors, like B, are not good choices for qubits, since the valence band degeneracy leads to rapid spin-lattice relaxation.

For control species, the needs are somewhat different, and the contrast in orbital extent between ground and excited states is especially important. Usually controls will have S=1/2, like substitutional P. If any other spin is used, the control is likely to leak quantum information. For S=0, for instance, excitation creates an electron (useful) and a hole, and the hole may relax its spin to cause decoherence. The control must have an accessible excited state that is reasonably delocalised, since this is what enables the qubit-qubit entanglement. Almost any diamond donor that has accessible excited states should do, even those unidentifed species that are detected in donor-acceptor pair transitions. However, some diamond Rydberg states not accessible, e.g., those that give rise to the GR2-GR8 lines of the neutral vacancy (Mainwood and Stoneham 1997). The control excited state should not have fast transitions to other states, since we want to use a stimulated absorption-stimulated emission sequence to control entanglements. This condition may well be met for substitutional P, but remains to be checked. Finally, the optical transition of any one control (with appropriate qubits) must be sharp in energy, so that many distinct gates can be resolved in the inhomogeneous line. The sharpness is an important factor in scalability.

We shall discuss architectures and characterisation later. Bearing in mind that a random distribution of controls and qubits will have some members that are placed so as to be useless, we might expect a minimal computer of say 10-20 gates to comprise some 100-200 dopants within a relatively small region, perhaps a few 100 nm across. There are relevant discussions of gate performance and its optimisation, and also the implementation of simple algorithms in Kerridge et al (2007), Gauger et al (2008) and Del Duce et al (2009).

5.     **Dopants and Defects in Diamond: Specific systems**

The spectroscopic data for defects in diamond (e.g., Bridges et al 1990, Field 1992) includes symmetries of the excited states in many cases, mainly through careful applied stress work (e.g., Davies and Hamer 1976). Yet very little indeed is known about the radial extent of excited states or even about excited state ionisation energies. In this section we summarise some basic information on four important defect species, combining theory and experiment to estimate key parameters.

We shall assume that, for practical purposes, the several species will be stable in the selected charge states. It is not entirely obvious that this is true. For instance (private communication, Dr Jonathan Goss) if P and N donors are present, it is probably stable to have $P^+$ and $N^-$, rather than $P^0$ and $N^0$. But, in good insulators like diamond, equilibrium may be very slow to establish, and effective stability over months or even years may be achievable. Certainly (private communication, Professor Ken Haenen) P donor and B acceptor spectra can be seen in the same sample. These considerations may influence a choice between N and $NV^-$ as qubit. We remark that it should be possible to choose C or N isotopes without a large extra cost.

### 5.1 The isolated substitutional nitrogen dopant $N_s$

Substitutional nitrogen has been widely studied with spin resonance, giving very precise values for hyperfine constants. The centre has <111> symmetry, not through a Jahn-Teller effect, but because of the single occupance of an antibonding N-C orbital. Incoherent reorientation between the four equivalent <111> orientations has been studied (Ammerlaan and Burgemeister 1981) and, at room temperature, spin lattice relaxation and reorientation are comparable. It may be that spin relaxation is marginally slower than reorientation. The long spin lattice relaxation time, of order 1 ms at room temperature, is encouraging for quantum applications. Whilst photoionisation of $N_s$ has been identified, there seem to be no interesting bound excited states or useful optical transitions.

### 5.2 The Nitrogen-Vacancy Centre $NV^-$

The nitrogen vacancy centre usually refers to $NV^-$, rather than $NV^0$. The $NV^0$ charge state is not normally accessed in the operations of interest here (Felton et al 2008). Remarkably, important pioneering studies relevant to quantum information processing were carried out by Glasbeek some 20 years ago (van Oort and Glasbeek 1992), before spin manipulations were associated with quantum computing. The $NV^-$ centre has a relatively compact S=1 ground state, and its negative charge means capture of another electron is not favoured.

The $NV^-$ line width can be very narrow (e.g., Shen et al 2008, Fuchs et al 2008, Davies 1974), so it is possible to observe shifts of the excitation energies of individual centres as they are tuned by electric or stress fields (Tamarat et al 2006). Davies (1974) discusses its temperature dependence. Thus random charged defects will give inhomogeneous broadening. Even spin zero charged defects (perhaps $P^+$ ionised donors, or or $PV^-$) can cause a useful spread of excitation energies ( for a survey of inhomogeneous broadening mechanisms, with quantitative estimates, see Stoneham 1969). Initialisation and readout of $NV^-$ centres have been demonstrated optically at room temperature (Hanson et al 2006).

Decoherence is relatively slow (Redman et al 1991, Kennedy et al 2003, Takahashi et al 2008, Jiang et al 2008), and has been studied and appears to be understood, mainly in terms of spin-spin and spin-lattice relaxation. Often S=1 defects seem less useful than S=1/2 ones, since $\delta S=1$ transitions tend to have faster spin lattice relaxation (decoherence) and bigger spin-spin relaxation rates; for $NV^-$, this does not seem a major issue.

Single $NV^-$ defects have been studied with optically-detected magnetic resonance at room temperature (Gruber et al 1997). Optical Rabi oscillations have been observed (Batalov et al 2008). Single electron spin readout has been demonstrated in a time that is shorter than the decoherence times by working at low temperatures (Jelezko et al 2002). Simple single centre (A-gate) operations have been carried out optically. Both nuclear and electron spin options for qubits have been exploited, recognising that both $^{13}C$ and $^{14}N$ spins can be important (Neumann et al 2008, Lovett and Benjamin 2009, Neumann et al 2009). The $^{14}N$ leads to hyperfine splittings (I = 1, hence $M_I$ = +1, 0, -1). Kennedy et al (2002) have suggested using the N spin and the NV$^-$ electron spin for a two-spin gate. This is an important achievement, but does not seem to offer a route to scaleability.

Spin dynamics has been studied in the excited state, and Autler-Townes behaviour, Hahn echoes, and other coherences have been observed (Tavares et al 1994; Jelezko et al 2004a, b; Batalov et al 2008).

**5.3 Other nitrogen centres related to radiation damage**

If we want qubit and control dopants for room temperature, then the ideal candidates would have a small Huang-Rhys factor (to ensure as much as possible of the intensity in the zero-phonon) and a very narrow zero-phonon line, even at these temperatures. We should note that the Huang Rhys factor depends on *first order* [linear] electron-phonon coupling, whereas zero phonon linewidths depend on *second* order couplings (Stoneham 1975 Chapter 14).

There are many centres that have sharp zero phonon lines at 77K, but far fewer still sharp at room temperature. Some such lines are discussed by Davies (1974), where there is an analysis of the mechanisms of broadening and line shifts with temperature. His analysis of seven centres (NV$^-$ at 1.945eV, a centre at 2.086 eV, the H3 centre at 2.463 eV, N3 at 2.985 ev, ND1 at 3.150ev, a centre at 3.188 ev, and the 5.251 ev N9 centre) implies that there are many centres in diamond with very sharp zero phonon lines at 77K. We are also indebted to Professor Alan Collins [private communication, 2008] for identifying some that are sharp at room temperatures, namely the N3 centre (2.985 eV, a vacancy with three N neighbours), the H2 centre (1.257 eV, an $N_sVN_s$ complex in an negative charge state), the H1b centre (0.6124 eV, a radiation complex including $N_s$-$N_i$) and the H1c centre (0.64 eV, a radiation complex involving N). It is notable that most of the interesting centres involve substitutional nitrogen with radiation damage defects like vacancies

**5.4     The isolated substitutional P donor**

**5.4.1  Spectroscopy**          The phosphorus donor in diamond is tantalising. A number of spectroscopic measurements (e.g., Nesladek et al 1999, Gheeraert et al 2000, Haenen et al 2001, Barjon et al 2007, Lazea et al 2008) show a centre whose optical transitions can be mapped onto those for an effective mass donor with ionisation energy of order 0.6 eV. Results for bound excitons (Lazea et al 2008) are consistent with these ideas, if one follows the empirical Haynes Rule.  This ionisation energy is large for traditional semiconductor applications, but in a very convenient range for some approaches to quantum computing, such as the SFG strategy. The P donor transitions broaden at room temperature, but it is not clear how much of the energy spread is inhomogeneous broadening and how much is phonon-induced homogeneous broadening.  One interesting conclusion (Jones et al 1996) is that the PV$^-$ centre has a spin zero ground state, and so cannot be exploited using the same tricks as the NV$^-$ centre. However, the charged PV$^-$ centres can contribute beneficially to inhomogeneous broadening.

Electron spin resonance measurements have established the g-factor and hyperfine constants of substitutional phosphorus centres introduced by chemical vapour deposition (Katagiri et al 2006).

What makes these ideas tantalising is the lack of information on the diamond conduction band effective masses, since it is not possible to separate out the central cell corrections. We shall try instead to put bounds on quantities relevant for quantum information processing. Such scoping calculations can be done using relatively simple methods, though fuller approaches will be needed once more experimental information is available.

**5.4.2  Theory for the ground state**    Several studies (Gheeraert et al 2000, Wang and Zunger 2002, Butorac and Mainwood 2008, Eyre et al 2005) have assessed the P donor in diamond, with emphasis on the ground state and its symmetry, and on the ionisation energy. The results are broadly in accord with observation. There is an important point (Dr J P Goss, private communication) that, if substitutional N and P are both present, it is probably energetically favourable for the P to donate its electron to the N, giving $P^+$ and $N^-$. The extent to which this happens in a highly insulating system needs an experimental check. Provided useable $P^0$ remain, as the charged defects have zero spin, their main effect will be to enhance inhomogeneous broadening, which could be desirable.

**5.4.3  Theory for excited states: effective mass wavefunctions**  There are remarkably few serious calculations of excited states of shallow defects in Group IV systems, and especially so for diamond. Experiment gives little information on one key aspect needed here, namely the extent of the excited state wavefunction in diamond. Even the electron effective mass is not known with useful accuracy for diamond, and all sensible values would imply a wavefunction towards the compact limit of effective mass theory. Nonetheless, effective mass theory often works well even close to its expected limits of validity, and we may use it with caution to scope important parameters. The results from the four simple models below give a reasonably consistent and certainly helpful estimate of the important interactions in the scoping calculations we give later.

*Method 1*   Suppose we know the effective Rydberg $R_{eff}$ (the ionisation energy, ideally after removing any central cell corrections). Then this should be the real Rydberg (13.6eV) multiplied by Rydberg $(m^*/m_0)/\varepsilon^2$, i.e., $(m^*/m_0)/\varepsilon^2 \sim R_{eff}/13.6eV$. The dielectric constant $\varepsilon = 5.7$ (Field 1992). The effective Bohr radius is $a_0 \varepsilon/(m^*/m_0)$, thus 0.529Å $[13.6/5.7]/R_{eff}(eV)$, giving 3.98 $a_0$ (or 2.1 Å) for $R_{eff}$=0.6 eV. However, there could be a central cell term that would give the wavefunction a greater spread. So, if the effective Rydberg without central cell were say 0.4 eV, then the effective Bohr radius would be larger by a factor 1.5 (3.16 Å). This sounds small, but the real question concerns the sizes of the interactions.

*Method 2*   Another route is to use the bound exciton information, since hole effective masses are known. To do this, however, one needs further information, such as the empirical Haynes Rule that the exciton binding is often about 0.1 of the donor binding (see Stoneham 1975 p 840). The bound exciton data are quite good, implying a binding of about 0.09 eV, hence - if we accept the 0.1 factor -  a donor energy of 0.9eV. This approach was used by Nakazawa et al (2001), who also used data for the boron acceptor ionisation energy, predicting 605 meV for the P ionisation energy.

*Method 3*   Other measures of the orbital extent in the ground state come from experimental and theoretical data as to interactions between donors. Thus one might analyse spin resonance data for systems like the N4 system to see when two N atoms become essentially undistinguishable from two separate N substitutionals. Goss (2008 Diamond Conference, unpublished) indicates that nitrogens at fifth neighbour separation have at least a small interaction. This is a relatively long range interaction, given that the very compact ground state of the N donor. Alternatively, one can look at optical spectra to check broadening by donor-donor interactions. From our own data (Dr Stephen Lynch, private communication, from data on samples provided by Professor K Haenen) we find that observed linewidths for P donors are broadly consistent with the estimates from methods 1, 2.

*Method 4*     Donor-acceptor pair spectra are observed in diamond (Dischler et al 1994), with the resolution of quite a few shells in emission (to shell 16 at 7.13 Å, then several more that may include one at 12.6 Å) and also in absorption (resolved to shell 13 at 6.37 Å). The resolution needed is of order perhaps 30 meV, i.e., not especially good. These results need a fairly deep donor (~3.2eV) so a shallow donor like P would give a more substantial range.

All these estimates suggest interactions between an excited control and a qubit over a useful range of spacings. Using these several estimates, we can get orders of magnitude for exchange interactions at various spacings, following the approach of Stoneham and Harker (1975a, b). Our results for diamond can also be compared with similar calculations for silicon.

## 6    Scoping potential for Quantum Information Processing

The gaps in knowledge about the spatial extents of electronic excited states of dopants in diamond - both experimental and theoretical - makes it hard to replicate the detailed calculations quantum gate dynamics that are possible for silicon (Kerridge et al 2007). At this stage, however, the main need is to scope the system, to see if there is a respectable chance that P- and N-doped diamond might work as a quantum information processor. Our emphasis is on the orders of magnitude of the key parameters for the SFG approach. We are, of course, well aware that defect and dopant engineering in diamond involves very substantial practical difficulties, so another concern is how severe these problems will be. We emphasise again that one important feature of the SFG approach is its use of a random distribution of dopants and its exploitation of inhomgeneity.

The calculations we do will be simple, based on minimal basis sets for the effective mass envelope wavefunctions. We emphasise that these are scoping calculations, not attempting the most sophisticated methods currently available. Technically, our approach is readily generalised, but such enhancements are not appropriate in a case like this. As known for some time (Shaffer and Williams 1964, Stoneham 1975 sections 4.2.7, 25.3, Andres et al 1981, Koiller et al 2002) there will also be a factor in any matrix elements from the band Bloch functions. This will give either enhancement or suppression, depending on precisely on the relative spacings of the sites involved, so certain combinations of positions may be ineffective. This is not a major problem, since any computer will exploit a selection of qubits and controls for which the interactions are shown to be appropriate.

In figure 2a, results are given for two P donors, taking values for a 0.6eV ionisation energy with no central cell corrections. Parallel calculations have shown that there is no change in overall predictions if we achieve the 0.6 eV ionisation energy with 0.4 eV Coulombic binding plus a central cell term. The blue curves are the 1s-1s exchange terms, and the red the 2p-1s exchange, i.e., with an excited electron. The energies are typically about ten times as large as for P in Si at the maxima, so there are still usable exchange energies out to 20 or 25 Angstroms. The likely qubits might be N or $NV^-$, for which a smaller effective radius is appropriate. Effective mass theory is not appropriate for either of these potential qubit defects, but we can see the impact of their more compact nature by reducing the effective Bohr radius of the qubit in the calculations (Figure 2b). Halving the qubit radius reduces the interaction by about a factor 2, but doesn not alter the distance dependence much, since that is determined primarily by the radius of the excited P donor. In the figures, we need the region in which the blue curve is much smaller than the red one, say beyond 18Å for 0.4eV or 9Å for 0.6eV. A sphere of radius 10Å centred on a lattice site contains 742 atom sites, one of radius 18 Å contains 4327 sites, and one of 25 Å contains 11592 sites. Thus a *local* concentration of even 100 ppm of the dopant might be effective. Dopant spacings of order say 2nm or less are needed. Within a region even 20 nm across, one might have the necessary few hundred dopants.

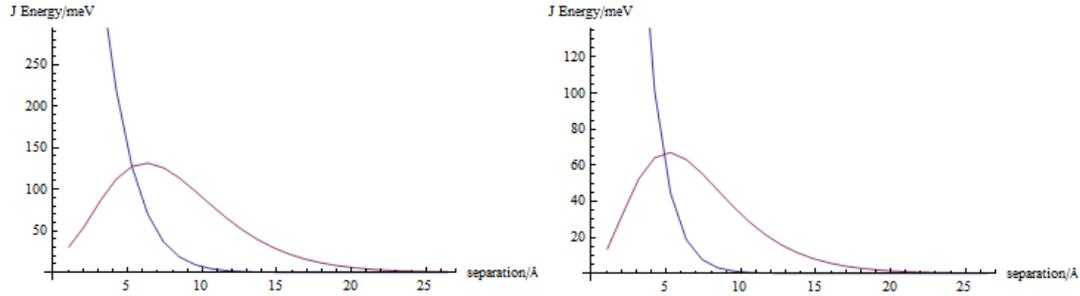

*Fig 2 (a) The exchange energy between two donors in diamond as a function of separation. The blue line is for both donors in their ground states, the purple line for one in the ground state and the other in the first excited P state. (b) The exchange energy between two different donors in diamond as a function of separation, one a control, as in (a), the other a more compact qubit with an effective Bohr radius half that of the control. The blue line is for both donors in their ground states, the purple line for qubit in its ground state and the control in its first excited P state.*

**7. Distributions of dopants, interactions and architectures**

**Architectures and Inhomogeneity** Assume we have a diamond in which there are $N_q$ potential qubits and $N_g$ potential gates in a region that can be addressed optically at chosen wavelengths, intensities and pulse lengths. The practical limit on the number of gates is determined by the number of separately resolved transitions in the inhomogeneously broadened absorption line of the control species. As an initial guess, we shall assume, for a minimal computer, both $N_q$ and $N_g$ will be of order 20 within any region whose size is resolvable optically, typically a few 100 nm across. There will, of course, be other dopant and defect species present, and these we assume play no part, apart perhaps from causing some of the (desirable) inhomogeneous broadening.

The spacings of the $N = N_q+N_g$ active species will be of the order of a few nm, consistent with the interactions discussed in section 4. So these N active species might be confined within a region of about 10 nm across, rather than spread through the optically-resolvable region. In other words, this minimal computer of 20 or so gates (qubits) could fit comfortably in a CVD grain or even in a nanodiamond (nanodiamond takes various meanings in the literature, even to a diamond 40µm across!). If a small diamond - a grain or a nanodiamond - is used, the surface will clearly have to be passivated to eliminate spins, and we must be aware that the proton spins of hydrogen, a common passivator, could cause some decoherence. However, the passivating protons form only a two-dimensional surface layer. There will be fewer protons than if they were spread through the bulk (the solubility of H in diamond is very low), and it should be possible to arrange that the active quantum dopants are not close to the surface.

The gates and qubits would occupy only a small fraction of any one optically-resolvable region. Irrespective of whether a microdiamond or a single crystal is used, a number of these "patches" of 20 or so qubits will have to be linked to form a serious computer. Since quantum computing power rises dramatically with qubit number (as $2^N$, at least of error correction is unnecessary) even a dozen of such patches - giving 200 or so qubits - would be a significant computer. But connecting patches is a challenge. There are various ideas, based on electrons (e.g., Stoneham 2008) or photons (perhaps exploiting the methods of Benjamin et al 2009). These will be analysed further in a paper in preparation.

**Statistical issues**   The assumption so far is that the distribution of dopants can be random, with nothing done to encourage special spacings. How efficient is this random doping?

Suppose we have randomly distributed dopants X (X=N, P). If we want there to be one dopant within the first five shells of neighbours, then the doping level must be high, over 2% atomic. If we chose to have 1% atomic, then the Poisson distribution tells us that 64.4% of dopants will have no other dopant within the first five shells, 28.3% will have one dopant, 6.2% will have two dopants, and only 1.1% have more than two dopants. This situation is not quite so bad as it sounds, since the J values may well be good enough out to say 10-20 Å, but it may be necessary to choose a region where the properties have proved suitable. Typical dopings of P donors give average concentrations of a few parts in $10^4$, which is still small; however, we need only a local region to have a raised concentration.

**Inhomogeneous broadening and its origins**   We need to be able to excite a chosen P (or, strictly, a chosen P/N/N group that constitute a gate), and here we need to combine (a) optical focus - spatial resolution - with (b) spectral resolution, exploiting the variation from one site to the next of energies due (i) to interactions between dopants and (b) due to say surface effects and other inhomogeneous broadening. The inhomogeneous shifts can come from the qubit energies, i.e., those of $NV^-$ or $N_s$ instead of just P. Overlap probably shifts the P peaks, whereas it may be electric fields or stress that shift the $NV^-$ line; such shifts have been quantified (Tamarat et al 2006). The fields could be from $N^+$ or $PV^-$ or other charged species.

Shen et al (2008) showed that $NV^-$ have linewidths nearly lifetime limited in 40 μm microdiamonds. In these small diamonds, the smallest linewidths seem to be about 0.36 nm, which is presumably largely homogeneous broadening $\delta_h$. The spectral distribution ranges over some 5nm, suggesting significant inhomogeneous broadening $\delta_i$. The ratio $\delta_i/\delta_h$ is thus probably of order 5/0.36 ~ 14, which suggests that perhaps 10 gates might be achieved, even with these unoptimised parameters. There are various ways that this ratio might be made larger. Relatively modest densities of charged defects, like $N^+$ or $PV^-$ would probably cause useful extra inhomogeneity, and raise the number of resolvable gates.

## 8   Configuring a diamond device: which wavelength operates which gate?

The *configuring* step establishes just which optical excitation energies actuate which gates. In an ideal world, this step would be done once only, and the device then operated many times. How might this be done, assuming that we have very sensitive, high resolution infrared (IR) and electron spin resonance (EPR) equipment that - by signal averaging over long periods - can have effective single centre detection? Ideally, we shall not need to know what the dopant positions are. In this section, we use what we know of the theory and energies to reassure ourselves that there will be acceptable resolution and information for configuration.

The configuring process would begin with a measurement of the EPR spectrum as a function of the optical excitation frequency. This yields a two-dimensional scan that provides the needed information about which optical transition affect which EPR transition and by how much. For N useful qubits, it is necessary to have N distinguishable (resolved) EPR resonances with controllable interactions. A qubit's interactions are controllable for our purposes if the qubit is initially not interacting with other qubits, but can be coupled to one or more qubits by an optical resonance that can be resolved from the other optical resonances that are already being used for coupling. In the simplest useful case, exciting a particular optical resonance would shift the position of exactly two EPR resonances, indicating that these two EPR spins can be coupled with such an optical frequency. In practice, some optical excitations would couple more than two qubits, which would not be a problem as long as the quantum algorithm was able to make use of this.

Further information would then be obtained by going beyond spectroscopy to pulsed resonance experiments. Higher power would be required for both the EPR and optical radiation so that Rabi oscillations can be recorded for each resonance to be used. Finally, one would apply a π pulse to each optical transition and measure how long the control should be left in the excited state to produce a particular entangling gate. Most quantum algorithms require just one such entangling gate, and here it is important to choose one which leaves the control unentangled with the qubits (Rodriquez et al 2004). The degree of entanglement between the qubits would be followed with pulsed EPR experiments similar to those used to violate Bell's inequalities (Neumann et al 2008, Lovett and Benjamin 2009, Neumann et al 2009).

The procedure described here is a gradual buildup towards doing a quantum computation. It appears that if a quantum computation can be done with a system where we can observe each resonance used, then we can configure the system by studying it. This hypothesis would equally apply to systems other than diamond.

Consider for illustration a system of $N_q$ qubits and $N_c$ controls. If $N_q$ were 3 and $N_c$ were 2, these would suffice for the minimal Deutsch-Josza case. Number the qubits 1, 2, … and controls A, B, etc. We want to know which qubits are controlled by which spins, and we want to do this from the EPR and IR data alone. For an initial analysis, some simplifying assumptions are appropriate. First, assume that the IR transitions for the controls are separated sufficiently in energy by inhomogeneous broadening. Secondly, suppose the EPR signals of controls and qubits are distinct. Thirdly, assume the EPR signals of the qubits are separated by inhomogeneous broadening. We may reasonably assume further that the qubit orbitals do not interact with each other or with controls in their ground states and, as a working assumption, that, in the excited state, the excited control electron interacts with only two of the three qubits.

Our scoping calculations can guide us as to whether these processes might be achieved. We can estimate all the key energies for the approach outlined above, following Stoneham and Harker (1975a,b) to obtain exchange and bonding-antibonding splittings. At this stage, we must now calculate operationally-relevant properties of a quantum device for *chosen* dopant positions and spacings. A choice is needed simply as input for theory; dopants do not need to be placed precisely in any real working device.

Figure 3 shows the energy dependence on separation for two symmetrical centres, using minimal basis set estimates for the two dopants involved. The limited basis set will lead to overestimates of interactions and energy shifts at shorter ranges; from comparisons with other calculations, spreads of order 30-40 meV are surely realistic. Calculations with larger basis sets are in hand. The corresponding exchange interactions can also be calculated (Fig 2a).

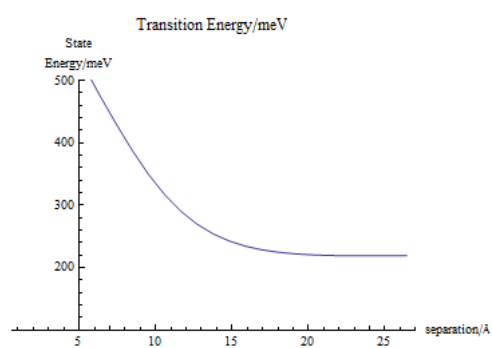

*Fig 3 Transition energy, meV, for two controls, with parameters for 0.6 eV Coulombic binding. The small basis set probably exaggerates the change in splitting at small separations.*

As illustration, take a small system of two controls and three qubits, sufficient for an extended Deutsch-Josza calculation. For realism, *low* symmetry is essential. The approach is are applicable to much larger systems subject to our computational limits. In our illustrative example, we take a two-dimensional system for simplicity, but our methods are not restricted in this way. Without loss of generality, we put the two controls symmetrically about the origin, C1 at (-a, 0) and the C2 at (a, 0) The three qubits Q1, Q2, Q3 we place at (-a-d, d/2), (-0.1a, d), (a, -d). We choose values of a, d that give sensible distances e.g., a = 12 Angstroms, d = 9 Angstroms. Given the chosen geometry, we may first calculate excitation energies. More precisely, we wish to confirm that the two control excitations will be resolvable. We can also calculate exchange interactions, since it is the qubit-excited state control exchange interactions that lead to the effective couplings between qubits that allow manipulation of entanglement.

Not surprisingly, the exchange interactions (Table 1) vary from substantial (C1 with Q1, Q2, and C2 with Q2, Q3) to negligible (C1-Q3, C2-Q1). In this particular case, each control only interacts significantly with two qubits; in a more general random distribution, more than two qubits might be coupled. Whilst one could also calculate qubit-qubit interactions, the present approach - at the limits of effective mass theory for the controls - would not be appropriate for the compact $N_s$ or $NV^-$ centres. Fortunately, there are far more detailed calculations (Dr J P Goss, private communication, 2008 Diamond Conference) that such interactions are very small beyond fifth neighbours. The values of control-qubit interactions are such that the configuring process outlined above should be possible. For manipulations of spins using the SFG approach, what we need to know as well are effective interactions between Q1, Q2 (gate 1) and between Q2, Q3 (gate 2). These are of order (CiQj exchange).(CiQk exchange)/(excitation energy), these very approximate estimates giving 0.7 meV for gate 1 and 0.4 meV for gate 2. Such values are in an acceptable range.

Table 1 Exchange interactions between the excited controls and qubits. For these calculations, the two controls are C1 at (-a, 0) and C2 at (a, 0), with a=10Å The three qubits Q1, Q2, Q3 are at (-a-d, d/2), (-0.1a, d), (a, -d) with d=9 Å.

| Control-Qubit | Separation (Angstrom) | Exchange, meV |
|---|---|---|
| C2 Q3 | 9 | 41.2 |
| C1 Q1 | 10 | 32.3 |
| C1 Q2 | 14 | 10.5 |
| C2 Q2 | 16 | 5.6 |
| C1 Q3 | 25.6 | 0.2 |
| C2 Q1 | 33.3 | 0.01 |

## 9     Conclusions

We have combined experimental data with scoping calculations of other quantities to assess whether one might make a modest quantum computer (perhaps 10-20 qubits) in diamond, following the approach of Stoneham, Fisher and Greenland. This The SFG approach looks promising for qubit manipulations, and decoherence (primarily spin-lattice relaxation, and spontaneous emission) appears to be acceptable. Readout and initialisation have already been demonstrated by others using the special properties of the $NV^-$ centres, and were not re-analysed here. However, we note sequential readout of qubits may not be the optimum approach. The configuring step (establishing which wavelengths control which gate) looks feasible for a system that is capable of acting as a computer.

We defer full discussions of two important topics to future papers. First, we shall examine elsewhere other options for readout. Secondly, we shall outline elsewhere ways that might allow the creation of a serious quantum computer by linking a number of mini-computers of 10-20 qubits. The standard idea is that one might use some form of "flying qubit", but proposals so far leave much to be desired. One other possible route is to find a way to inegrate the mini-computers with a measurement-based approach to quantum computing (Benjamin et al 2009). Measurement-based QIP exploits the generation of quantum entanglement between remote systems by performing measurements on them in a certain way. The systems might be two of the mini-computers, each containing a single $NV^-$ centre prepared in specific electron spin states, the two centres tuned to have exactly the same optical energies. The measurement involves 'single shot' optical excitation., exposing both systems to a weak laser pulse that, on average, will achieve one excitation. The single system excited will emit a photon that, after passing though beam splitters and an interferometer, is detected without giving information as to which system was excited. 'Remote entanglement' is achieved, subject to some strong conditions. For effective links between mini-computers, the strategy needs to be changed, but there is here an important feature that links can be made over quite large distances. All this is still very demanding, and might still be defeated by issues of system integration. But it seems optimistic enough to go forward.

**Acknowledgments**  This work was supported by RCUK and EPSRC through the Basic Technologies programme. We are indebted to Professor Gabriel Aeppli, Dr Simon Benjamin, Dr Jonathan Goss, Professor Ken Haenen, Dr Brendon Lovett, Dr Stephen Lynch and Dr Thornton Greenland for valuable discussions.

**Figure captions**

FIGURE 1: The Stoneham Fisher Greenland (SFG) approach. Quantum gates exploit optical control of electron spins. Coupling between qubits (green dots) is only significant when the control dopants (red dots) are excited. The laser source can focus only to about a square wavelength, far larger than spacings of a few nm needed for entanglement. Selection of a particular gate (control plus qubits) exploits inhomogeneous broadening: different wavelengths excite different gates in a random system. Thus in (a) red entangles one group, whereas (b) green entangles another group, and leaves the first group unaffected; (c) after two excitations, two gates have made controlled changes in entanglements

Fig 2 (a) The exchange energy between two donors in diamond as a function of separation. The blue line is for both donors in their ground states, the purple line for one in the ground state and the other in the first excited P state. (b) The exchange energy between two different donors in diamond as a function of separation, one a control, as in (a), the other a more compact qubit with an effective Bohr radius half that of the control. The blue line is for both donors in their ground states, the purple line for qubit in its ground state and the control in its first excited P state.

Fig 3 Transition energy, meV, for two controls, with parameters for 0.6 eV Coulombic binding. The small basis set probably exaggerates the change in splitting at small separations.

Table 1 Exchange interactions between the excited controls and qubits. For these calculations, the two controls are C1 at (-a, 0) and C2 at (a, 0), with a=10Å The three qubits Q1, Q2, Q3 are at (-a-d, d/2), (-0.1a, d), (a, -d) with d=9 Å.

| Control - Qubit | Separation (Angstrom) | Exchange, meV |
|---|---|---|
| C2 Q3 | 9 | 41.2 |
| C1 Q1 | 10 | 32.3 |
| C1 Q2 | 14 | 10.5 |
| C2 Q2 | 16 | 5.6 |
| C1 Q3 | 25.6 | 0.2 |
| C2 Q1 | 33.3 | 0.01 |